\documentclass[vecphys]{svmult}

\usepackage{makeidx}         
\usepackage{graphicx}        
\usepackage{multicol}        
\usepackage[bottom]{footmisc}

\makeindex             

\begin{document}

\title*{Gravitational stability in the disk of M51}

   \author{  
     M.\,Hitschfeld\inst{1} \and  
     C.\,Kramer\inst{1} \and  
     K.F.\,Schuster\inst{2} \and  
     S.\,Garcia-Burillo\inst{3} \and  
     J.\,Stutzki\inst{1} 
          }

   \institute{  
     KOSMA, I. Physikalisches Institut, Universit\"at zu K\"oln,  
     Z\"ulpicher Stra\ss{}e 77, 50937 K\"oln, Germany \\  
     \texttt{hitschfeld@ph1.uni-koeln.de}, \texttt{kramer@ph1.uni-koeln.de},\\
     \texttt{stutzki@ph1.uni-koeln.de}
     \and  
     IRAM, 300 Rue de la Piscine, F-38406 S$^t$ Martin d'H\`{e}res, France  \\
     \texttt{schuster@iram.fr}
     \and  
     Observatorio de Madrid, Alfonso XII, 3, 28014 Madrid, Spain        \\
     \texttt{s.gburillo@oan.es}
    }

\maketitle

\section{Abstract}

Star formation laws, like i.e. the Schmidt law relating star formation rate
and total gas density , have been studied in several spiral galaxies but the
underlying physics are not yet well understood.\\
M51, as a nearby face-on galaxy grand design spiral studied in many line transitions, is an
ideal target to study the connection between physical conditions of the gas and star
formation activity.
In this contribution we combine molecular, atomic, total gas and stellar surface densities and study the
gravitational stability of the gas (Schuster et al.2007, Hitschfeld et al. in prep.). \\
From our IRAM-30m $^{12}$CO2-1 map and complementary HI-, Radio Continuum- and ACS B-band-data
we derive maps of the total gas density and the stellar surface density to
study the gravitational stability of the gas via the Toomre Q parameter. As an important factor in this analysis we also
present a map indicating the velocity dispersion of the molecular gas
estimated from the equivalent widths $\Delta v_{\rm eq}$ of the $^{12}$CO2-1 data.

\section{The velocity dispersion of the molecular gas}
The velocity dispersion is important for the calculation of the  Toomre
parameter as it is hindering gravitational collapse. 
The map of the equivalent widths of  $^{12}$CO 2--1  $\Delta v_{\rm eq}=\int T
dv/ T_{\rm pk}$ is shown in Fig.1. 
It is related to the velocity dispersion via $\sigma_{\rm CO} = \Delta v_{\rm eq}/(2\,\sqrt{2\,\ln2})$.\\
The equivalent widths drop from the center to the outskirts by up to a
factor of 5 from  less than $\sim$ 20 kms$^{-1}$ to 102 kms$^{-1}$. The inner spiral arm structure of M51 in the northern
part is much less prominent than in the integrated intensity
 map (Schuster et al. 2007). 
\begin{figure}[h]
\centering
\includegraphics[height=4.5cm,angle=-90]{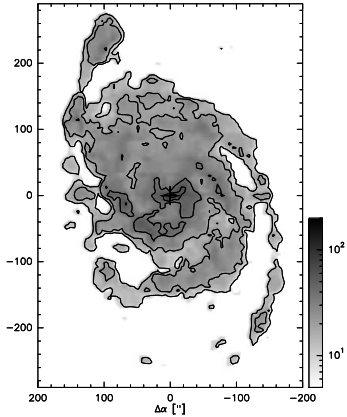}
\caption{The map of the equivalent widths of  $^{12}$CO 2--1  $\Delta v_{\rm
    eq}$. Contours show 10, 20, 40, 60 to 80 kms$^{-1}$.}
\end{figure}
\section{The Toomre Q-parameter}
The Toomre Q-parameter (Toomre 1964) describes the instability of a differentially
rotating, homogeneous thin gas disk against axial symmetric disturbances.
It is related to the epicyclic frequency $\kappa$, the velocity dispersion
of the gas $\sigma_{\rm gas}$ and the total gas surface density $\Sigma_{\rm
  gas}$ via:
\begin{equation}  
Q_{gas} = \frac{\kappa \sigma_{\rm gas}}{\pi G \Sigma_{\rm gas}}.  
\label{eq_sigma_crit}
\end{equation}
The epicyclic frequency is determined from the rotation curve of M51, we
assume that  $\sigma_{\rm gas}$ can be estimated from $\sigma_{\rm CO}$. The
surface density $\Sigma_{\rm gas}$ is constructed from complementary VLA-HI
data and the $^{12}$CO2-1 data as described in Schuster et al. 2007.\\
As a next step of the stability analysis the stellar component is taken
into account (Hitschfeld et al. 2007). The Q-parameter for a pure stellar disk
takes the equivalent form as Q$_{gas}$. The epicyclic frequency determined
from the rotation curve is identical. A good approximation for an combined
Q-parameter (Wang\&Silk 1994) is:
\begin{equation}  
Q_{total}^{-1} = Q_{gas}^{-1}+Q_{*}^{-1}.  
\label{eq_sigma_crit}
\end{equation}  
To calculate $Q_{*}$ we obtained the HST-ACS B-band image of Mutchler et
al. (2005) and converted it to a stellar mass surface density assuming a 
constant mass-to-luminosity ratio $M_{*}/L_{B}=1.54$ (Shetty et al. 2006).\\
The stellar velocity dispersion is significantly larger than the gas velocity
dispersion and can be estimated using an exponential fall-off (Botema et
al. 1993) depending on the stellar scale height of the disk.


\subsection{Results}

\begin{figure}[h]
\centering
\includegraphics[height=7cm,angle=-90]{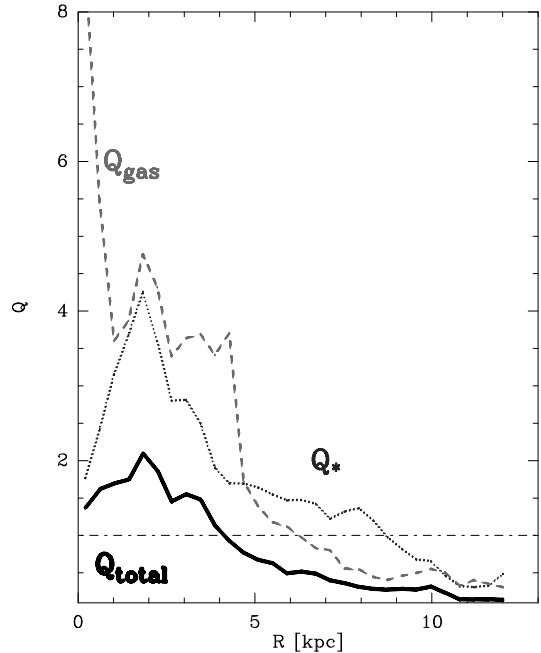}
\caption{Radial averages of the Toomre Q-parameter for the total gas $Q_{gas}$, stellar
  component $Q_{*}$ and the total $Q_{total}$. All parameters have been calculated from 
  inclination corrected quantities. A horizontal line delineates Q=1 for
  critically stable conditions.}
\end{figure}

The radial averages of Q$_{*}$,Q$_{gas}$ and Q$_{total}$ summed in elliptical annulli of 10" 
are presented in Fig.2. The importance of the stellar
contribution is evident and lowers the Q-parameter by up to 50\%. Due to the
stars the inner part is significantly closer to gravitational collapse than
estimations from the gas-only analysis would predict. The gravitational
collapse and formation of Giant Molecular Clouds is thus possible and
predicted over large regions of the disk. This agrees with previous studies by
Boissier et al.(2003) of radial averages of the Toomre
Q-parameter in their sample of galaxies comparing the gas-only $Q_{gas}$ and $Q_{total}$.\\ 
\begin{figure}[h]
  \centering
 \includegraphics[height=5.5cm,angle=-90]{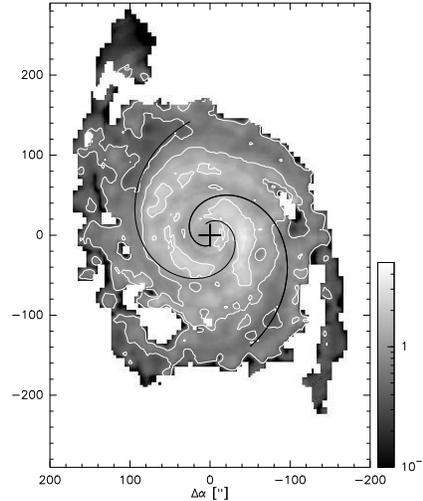}   
 \caption{Map of the combined Toomre-parameter of the stellar and
  gaseous component Q$_{total}$. In black two logarithmic spirals indicate the
  position of the inner spiral arms.} 
 \end{figure}
The map of the combined Q-parameter is shown in Fig.3.
The total Q exceeds 1 in the inner interarm regions and drops to beeing
critical to collapse (Q$\le$1) in the spiral arms indicated by the two black logarithmic
spirals. From galactocentric radii of around 4 kpc and beyond Q drops below 1 indicating
gravitational instability. Note that the Q parameter in the neighbour galaxy
NGC5195 is significantly underestimated due to the exponential decline assumed
for the velocity dispersion of the stars.\\
Star formation is observed over large scales in the disk of M51 and in particular 
massive, young stars are found in the inner part and especially in the spiral
arms. Since not only gravitational stability governs star formation but also
mechanism like i.e. spiral density waves and more locally turbulence and
supernovae explosions, the Toomre criterion naturally fails to explain the
exact location of stars. It is important as a threshold and necessary
pre-condition for star formation.    
%
%

%
%



\printindex
\end{document}